\documentstyle[aps,12pt,manuscript]{revtex}
\begin{document}
\preprint{INJE-TP-98-2/hep-th9803080}

\title{Dynamical Behavior of the BTZ Black Hole}

\author{ H.W. Lee, N.J. Kim and Y. S. Myung}
\address{Department of Physics, Inje University, Kimhae 621-749, Korea} 

\maketitle

\begin{abstract}
We study the dynamical behavior of the BTZ(Banados-Teitelboim-Zanelli) 
black hole with the 
low-energy string effective action.
The perturbation analysis around the BTZ black hole 
reveals a mixing between the dilaton and other fields.  Introducing the 
new gauge(dilaton gauge), we disentangle this mixing completely and obtain 
one decoupled dilaton equation.  
We obtain the decay rate $\Gamma$ of BTZ black hole.
\end{abstract}

\newpage
The D-branes techniques were used to derive the Bekenstein-Hawking 
entropy for the extremal or near-extremal 4d and 5d black holes\cite{Str96}. 
On the other hand, Carlip's approach can be applied to the 
non-extremal black holes, but it seems to be suitable for 3d ones\cite{Car95}. 
Recently Sfetsos and Skenderis showed that 
4d black hole(5d black hole) correspond to U-dual to 
BTZ$\times S^2$(BTZ$\times S^3$)\cite{Sfe97}. They calculated the entropies 
of non-extremal 4d and 5d black holes by applying Carlip's 
approach to the BTZ black hole. The BTZ black hole(locally, anti-de Sitter 
spacetime:AdS$_3$) is actually an exact solution of string theory
\cite{Ban92,Hor93}. 
And there is an exact conformal field theory with it on the 
boundary. Carlip has shown that the physical boundary 
degrees of freedom account for the Bekenstein-Hawking 
entropy of the BTZ black hole correctly.

In this letter we investigate the 
dynamical behavior(absorption cross-section=greybody factor) of 
the BTZ black hole rather than the static 
behavior(entropy)\cite{Dha96,Cal97}.  Apart from counting the microstates 
of black holes, the dynamical behavior is also an important issue. This is 
so because the greybody factor for the black hole arises as a consequence 
of scattering off the gravitational potential barrier surrounding the horizon.  
That is, this is an effect of spacetime curvature. Together with the 
Bekenstein-Hawking entropy, this seems to be the strong hint of a deep 
and mysterious connection between curvature and statistical mechanics.
It was shown that the greybody factor for the BTZ black hole has the 
same form as the one for 5d black hole in the dilute gas approximation
\cite{Bir97}. 
In this case a minimally coupled scalar was used for calculation. 
However, 
fixed scalars play the important role in testing the dynamical 
behaviors.  Due to the non-minimal couplings, 
the low-energy greybody factors for fixed scalars are suppressed compared 
to those of minimally coupled scalars\cite{Cal97}.  Here the dilaton($\Phi$) 
is introduced 
as a fixed scalar to calculate the decay rate of the BTZ black hole. 

We start with the low-energy string action in string frame\cite{Hor93}
\begin{equation}
S_{l-e} = \int d^3 x \sqrt{-g} e^{\Phi}
   \big \{ R + (\nabla \Phi)^2 + {8 \over k} - {1 \over 12} H^2  \big \},
\label{action}
\end{equation}
where $\Phi$ is the dilaton, $H_{\mu\nu\rho}=3 \partial_{[\mu}B_{\nu\rho]}$ 
is the Kalb-Ramond field, and $k$ the cosmological constant.
The equations of motion lead to
\begin{eqnarray}
R_{\mu\nu} - \nabla_\mu \nabla_\nu \Phi 
-{1\over 4} H_{\mu \rho \sigma} H_\nu^{\rho \sigma} &=& 0,
\label{eq_graviton} \\
\nabla^2 \Phi + (\nabla \Phi)^2 - {8 \over k} - {1 \over 6} H^2   &=& 0,
\label{eq_scalar1}  \\
 \nabla_\mu H^{\mu \nu \rho} + (\nabla_\mu \Phi) H^{\mu \nu \rho} &=& 0.
\label{eq_anti}
\end{eqnarray}
The BTZ black hole solution to 
(\ref{eq_graviton})-(\ref{eq_anti}) is found to be\cite{Ban92}
\begin{eqnarray}
&& \bar H_{txr} = 2r/l, ~~~~~~\bar 
\Phi = 0,~~~~~~k=2 l^2,
      \nonumber   \\
&& \bar g_{\mu\nu} =
 \left(  \begin{array}{ccc}  (M - {r^2 / l^2}) & -{J / 2} & 0  \\
                             -{J / 2} & r^2 & 0  \\
    0 & 0 & f^{-2}
         \end{array}
 \right)
\label{bck_metric}
\end{eqnarray}
with $f^2 =r^2 / l^2 -M +J^2 / 4 r^2$.
The metric $\bar g_{\mu\nu}$ is singular at $r=r_{\pm}$,
\begin{equation}
r_{\pm}^2 = {{Ml^2} \over 2} \left \{ 1 \pm \left [ 
   1 - \left ( {J \over Ml} \right )^2 \right ]^{1/2} \right \}
\label{horizon}
\end{equation}
with $M=(r_+^2 + r_-^2) / l^2, J=2 r_+r_- / l$.
For convenience, we list the Hawking temperature $T_H$, the area of 
horizon ${\cal A}_H$, and the angular velocity at the horizon
$\Omega_H$ as
\begin{equation}
T_H = (r_+^2 - r_-^2) / 2 \pi l^2 r_+,
~~{\cal A}_H = 2 \pi r_+,
~~\Omega_H = J / 2 r_+^2.
\label{temp}
\end{equation}

To study the propagation specifically, we introduce 
the small perturbation fields 
($H_{txr} = \bar H_{txr} + {\cal H}_{txr}, 
\Phi = 0 + \phi, 
g_{\mu\nu} = \bar g_{\mu\nu} + h_{\mu\nu}$) 
around the background solution 
(\ref{bck_metric}) as \cite{Sfe92}.
For convenience, 
we introduce the notation
${\hat h}_{\mu\nu} = h_{\mu\nu}- {\bar g}_{\mu\nu} h/2$ with
$h= h^\rho_{~\rho}$.
And then one needs to linearize (\ref{eq_graviton})-(\ref{eq_anti}) 
to obtain
\begin{eqnarray}
  \delta R_{\mu\nu} (h) 
- \bar \nabla_\mu \bar \nabla_\nu \phi   
- {1 \over 2} \bar H_{\mu \rho \sigma} {\cal H}_\nu^{~ \rho \sigma}   
+ {1 \over 2} \bar H_{\mu \rho \sigma} \bar H_{\nu\alpha}^{~~\sigma} 
h^{\rho \alpha} &=& 0,
\label{lin_graviton} \\
 \bar \nabla^2 \phi 
- {1 \over 6} \Big \{ 2 \bar H_{\mu \rho \sigma} 
                      {\cal H}^{\mu \rho \sigma}                
     - 3 \bar H_{\mu \rho \sigma}   \bar H^{\alpha \rho \sigma} h^\mu_\alpha 
                                    \Big \} &=& 0, 
\label{lin_scalar} \\
   \bar \nabla_\mu  {\cal H}^{\mu \nu \rho} 
- ( \bar \nabla_\mu h_\beta^{~\nu}) {\bar H}^{\mu\beta\rho}
+ (\bar \nabla_\mu h_\beta^{~\rho}) {\bar H}^{\mu\beta\nu}
- (\bar \nabla_\mu {\hat h}_{~\alpha}^{\mu}) {\bar H}^{\alpha\nu\rho}
      + (\partial_\mu \phi) \bar H^{\mu \nu \rho}
 &=& 0, 
\label{lin_anti}
\end{eqnarray}
where the Lichnerowicz operator $\delta R_{\mu\nu}(h)$ 
is given by \cite{Gre94}
\begin{eqnarray}
&&\delta R_{\mu\nu} = -{1 \over 2} \bar \nabla^2 h_{\mu\nu} 
 +{\bar R}_{\sigma ( \nu} h^\sigma_{~\mu )}
 -{\bar R}_{\sigma \mu\rho\nu} h^{\sigma\rho}
 + \bar \nabla_{( \nu} \bar \nabla_{|\rho|} {\hat h}^\rho_{~\mu)}.
\label{delR}    
\end{eqnarray}
These are the bare perturbation equations.  We have to examine whether 
there exist any choice of gauge which can simplify 
(\ref{lin_graviton})-(\ref{lin_anti}).  
A symmetric 
traceless tensor has D(D+1)/2--1 in D-dimensions.  D of them 
are eliminated by the gauge condition.  Also D--1 are 
eliminated from our freedom to take 
further residual gauge transformations.  
Thus gravitational degrees of freedom are D(D+1)/2--1--D--(D--1)=D(D--3)/2.  
In three dimensions we have no 
propagating degrees of freedom for $h_{\mu\nu}$.  
Also $B_{\mu\nu}$ has no physical degrees of freedom for D=3.
Hence the physical degree of freedom in the BTZ black hole turns out to be 
the dilaton field.  

Considering the $t$ and $x$-translational symmetries of the background 
spacetime (\ref{bck_metric}),  
we can decompose $h_{\mu\nu}$ into frequency modes in these 
variables \cite{Gre94} 
\begin{equation}
h_{\mu\nu}(t,x,r) = e^{-i \omega t} e^{i \mu x}H_{\mu\nu}(r).
\label{ptr_metric}
\end{equation}
Similarly, one chooses the perturbations for Kalb-Ramond 
field and dilaton as
\begin{eqnarray}
{\cal H}_{txr}(t,x,r) &&= \bar H_{txr} {\cal H}(t,x,r) 
 =\bar H_{txr} e^{-i \omega t} e^{i \mu x} \tilde{\cal H}(r),
\label{ptr_anti} \\
\phi(t,x,r)&&=e^{-i \omega t} e^{i \mu x} \tilde\phi(r).
\label{ptr_scalar}
\end{eqnarray}
Since the dilaton is a propagating mode, hereafter we are 
interested in the dilaton equation (\ref{lin_scalar}). 
Eq.(\ref{lin_graviton}) is irrelevant to our analysis, because 
it belongs to the redundant relation. 
Eq.(\ref{lin_scalar}) can be rewritten as 
\begin{eqnarray}
\bar \nabla^2 \phi
-{4 \over l^2} (h - 2 {\cal H} )
=0.
\label{eq_scalar}
\end{eqnarray}
If we start with full degrees of freedom (\ref{ptr_metric}), we 
should choose a gauge.
Conventionally, we choose the 
harmonic(transverse) gauge($\bar \nabla_\mu {\hat h}^{\mu\rho} = 0$) to
describe the propagation of gravitons\cite{Wei72}.  
It turns out that a mixing between the dilaton and other fields is 
not disentangled with the 
harmonic gauge condition.  But if we introduce the 
dilaton gauge( 
$h^{\mu \nu} \Gamma^\rho_{\mu \nu} = 
\bar\nabla_\mu \hat h^{\mu \rho}$), 
the difficulty can be resolved\cite{Lee98}.
Now we attempt to disentangle the last term in (\ref{eq_scalar}) by 
using both the dilaton gauge and 
Kalb-Ramond equation (\ref{lin_anti}).
Each component of dilaton gauge condition gives rise to 
\begin{eqnarray}
t&:& (\partial_r + {1 \over r} ) h^{tr} 
- i \omega h^{tt} + i \mu h^{tx}  
+ {1 \over 2} i \omega h g^{tt} - {1 \over 2}i \mu h g^{tx} = 0, 
\label{eq_gauge_t}  \\
x&:& (\partial_r + {1 \over r} ) h^{xr} 
- i \omega h^{xt} + i \mu h^{xx}  
+ {1 \over 2}i \omega h g^{xt} - {1 \over 2}i \mu h g^{xx} = 0, 
\label{eq_gauge_x}  \\
r&:& (\partial_r + {1 \over r} ) h^{rr} 
- i \omega h^{rt} + i \mu h^{rx}  
- {1 \over 2} (\partial_r h) g^{rr} = 0. 
\label{eq_gauge_r}
\end{eqnarray}
And the Kalb-Ramond equation (\ref{lin_anti}) leads to 
\begin{eqnarray}
tx:&& -\partial_r (\phi+{\cal H} -{h^t_{~t}} -{h^x_{~x}})
+ {1 \over rf^2}\left (M -{3 r^2 \over l^2} + 
                     {J^2 \over 4r^2}\right ) h^r_{~r} 
+i\omega h^t_{~r} -i\mu h^x_{~r}=0,
\label{eq_anti_tx} \\
tr:&& -i\mu (\phi+{\cal H} -h^t_{~t} -h^r_{~r})
- {1 \over r} h^r_{~x} 
+2 f^2 h^x_{~r}
-\partial_r h^r_{~x} +i\omega h^t_{~x} =0,
\label{eq_anti_tr} \\
xr:&& -i\omega (\phi+{\cal H} -h^x_{~x} -h^r_{~r})
           + {1 \over r} h^r_{~t} 
+{2 rf^2 \over l^2} h^t_{~r}
+\partial_rh^r_{~t} +i\mu h^x_{~t} =0.
\label{eq_anti_xr}
\end{eqnarray}
Solving six equations (\ref{eq_gauge_t})-(\ref{eq_anti_xr}), one finds an 
important equation
\begin{equation}
\partial_\mu (2 \phi +2 {\cal H} - h ) =0, ~~\mu=t,x,r
\label{eq_anti_simple}
\end{equation}
which leads to 
$h - 2 {\cal H} =2 \phi$.
This means that $h-2 {\cal H}$ is a redundant field.  
Hence (\ref{eq_scalar}) becomes a decoupled dilaton equation
\begin{eqnarray}
\left [ f^2 \partial_r^2 
+ \left\{ {1 \over r} (\partial_r rf^2) \right\} \partial_r
-{{J \mu \omega} \over {r^2 f^2}} 
+{\omega^2 \over f^2} 
+{{M-{r^2 \over l^2}} \over r^2 f^2} \mu^2 
\right ] \tilde \phi
-{8 \over l^2}\tilde \phi 
=0,
\label{eq_decoupled}
\end{eqnarray}
Here from equation (\ref{eq_anti_simple}), a constant of 
integration may exist. However, this is not relevant to the physics. 
For example, let us suppose the differential equation 
(\ref{eq_decoupled}) with a constant.  By requiring two physical 
boundary conditions for $\tilde \phi$ at $r=r_+, \infty$, we can 
determine to be zero. 
It is noted that if the last term is absent, (\ref{eq_decoupled}) 
corresponds to the minimally coupled scalar.

We are now in a position to calculate the absorption cross-section 
to study the dynamical behavior of 
the BTZ black hole.  
Since it is hard to find a 
solution to (\ref{eq_decoupled}) directly, 
we use a matching procedure. 
The spacetime is divided into two regions: the near region ($r \sim r_+$) 
and far region ($r \to \infty$)\cite{Dha96,Cal97}.  
We now study each region in turn.
For the far region($r \to \infty$), the dilaton equation (\ref{eq_decoupled}) 
becomes
\begin{equation}
\tilde \phi_\infty'' + {3 \over r} \tilde \phi_\infty' 
+ {s \over r^2} \tilde \phi_\infty=0.
\label{eq_far}
\end{equation}
Here we introduce $s=-8-\epsilon$ with the small parameter $\epsilon$ for 
the technical reason.
First we find the far region solution
\begin{equation}
\tilde \phi_{\rm far}(r) = 
{1 \over x} \left ( \alpha x^{\sqrt{1-s}} 
                  + \beta x^{-\sqrt{1-s}} \right )
\label{sol_far}
\end{equation}
with two unknown coefficients $\alpha$, $\beta$ and $x=r/l$.
We need the ingoing flux at infinity and this is given by
\begin{displaymath}
{\cal F}_{\rm in}(\infty) = -2 \pi \sqrt{1-s} \vert \alpha - i \beta \vert^2.
\nonumber
\end{displaymath}

In order to obtain the near region behavior,
we introduce the variable 
$z={{r^2 - r_+^2} \over {r^2 - r_-^2}}=
{{x^2 - x_+^2} \over {x^2 - x_-^2}},~~0 \le z \le 1$.
Then (\ref{eq_decoupled}) becomes
\begin{equation}
z(1-z) {{d^2 \tilde \phi} \over dz^2}
+(1-z) {{d \tilde \phi} \over d z}
+\left ( {A_1 \over z} +{{s/4} \over 1-z} +B_1 \right ) \tilde \phi
=0,
\label{eq_hyper}
\end{equation}
where 
$A_1 = 
\left ( {{\omega - \mu \Omega_H} \over 4 \pi T_H} \right )^2, 
B_1 = - {r_-^2 \over r_+^2}
\left ({{\omega - \mu \Omega_H r_+^2 / r_-^2} \over 
4 \pi T_H} \right )^2$.
The solution for (\ref{eq_hyper}) is given by
\begin{eqnarray}
\tilde \phi_{\rm near}(z) &=&
C_1 z^{-i \sqrt{A_1}} (1 -z )^{(1 - \sqrt{1-s})/2} F(a,b,c;z)
\nonumber \\
&&~~~~~~~~+C_2 z^{i \sqrt{A_1}} (1 -z )^{(1 - \sqrt{1-s})/2} F(b-c+1,a-c+1,2-c;z),
\label{sol_hyper}
\end{eqnarray}
where
\begin{eqnarray}
a&=& \sqrt{B_1} - i \sqrt{A_1} +(1 - \sqrt{1-s} )/2,
\nonumber \\
b&=& - \sqrt{B_1} - i \sqrt{A_1} + (1 - \sqrt{1-s} )/2,
\nonumber \\
c&=& 1 - 2 i \sqrt{A_1}.
\nonumber 
\end{eqnarray}
and $C_1$ and $C_2$ are to-be-determined constants.
At the near horizon($r\sim r_+, z \sim 0$) (\ref{sol_hyper})
becomes
\begin{eqnarray}
\tilde \phi_{\rm near}(0) &\simeq& C_1 z^{-i \sqrt{A_1}} + C_2 z^{i \sqrt{A_1}}
\nonumber \\
&=&C_1 \left ( { 2 x_+ \over {x_+^2- x_-^2}} \right ) ^{-i \sqrt{A_1}}
           e^{-i \sqrt{A_1} \ln(x-x_+)}
+C_2 \left ( { 2 x_+ \over {x_+^2- x_-^2}} \right ) ^{i \sqrt{A_1}}
           e^{i \sqrt{A_1} \ln(x-x_+)}.
\label{sol_z0}
\end{eqnarray}
Considering an ingoing mode at horizon, we have $C_2=0$. 
Hence the near region solution is
\begin{eqnarray}
\tilde \phi_{\rm near}(z) &=&
C_1 z^{-i \sqrt{A_1}} (1 -z )^{(1 - \sqrt{1-s})/2} F(a,b,c;z).
\label{sol_near}
\end{eqnarray}
Now we need to match the far region solution (\ref{sol_far}) to the large 
$r(z\to 1$) limit  of near region solution (\ref{sol_near}) in the 
overlapping region.  The $z\to 1$ behavior of (\ref{sol_near}) follows 
from the $z \to 1-z$ transformation rule for hypergeometric functions. 
Using $1-z \sim (x_+^2 - x_-^2)/x^2$ for $r \to \infty$, 
this takes the form
\begin{eqnarray}
\tilde \phi_{n\to f}(r) &\simeq&
C_1 E_1
{x^{ \sqrt{1-s}}\over x} 
+C_1 E_2 
{x^{-\sqrt{1-s}}\over x},
\label{sol_z}
\end{eqnarray}
where
\begin{eqnarray}
E_1 &=& 
{{ \Gamma(1 -2 i \sqrt{A_1}) \Gamma(\sqrt{1-s})
    (x_+^2- x_-^2)^{(1-\sqrt{1-s})/2}} \over
 {\Gamma({{1+\sqrt{1-s}} \over 2} + \sqrt{B_1} -i \sqrt{A_1} ))
  \Gamma({{1+\sqrt{1-s}} \over 2} - \sqrt{B_1} -i \sqrt{A_1} ))}},
\label{ain} \\
E_2 &=&
{{ \Gamma(1 -2 i \sqrt{A_1}) \Gamma(- \sqrt{1-s})
   (x_+^2-x_-^2)^{(1+ \sqrt{1-s})/2}} \over
  {\Gamma({{1-\sqrt{1-s}} \over 2} + \sqrt{B_1} -i \sqrt{A_1} ))
   \Gamma({{1-\sqrt{1-s}} \over 2} - \sqrt{B_1} -i \sqrt{A_1} ))}}.
\label{aout}
\end{eqnarray}
Matching (\ref{sol_far}) with  (\ref{sol_z}) 
leads to $\alpha = C_1 E_1$ and $\beta = C_1 E_2$.  Assuming 
$x_+^2-x_-^2 \ll 1 $, $\beta \ll \alpha$.  The ingoing flux across 
the horizon is 
${\cal F}_{\rm in}(0) = -8 \pi \sqrt{A_1} (x_+^2- x_-^2) \vert C_1 \vert^2$.
Hence for $\mu=0$, we can obtain the absorption coefficient 
\begin{eqnarray}
{\cal A} &=& 
{{\cal F}_{\rm in}(0) \over {\cal F}_{\rm in}(\infty)} = 
{ 4 \sqrt{A_1} (x_+^2-x_-^2) \over \sqrt{1-s} }
{1 \over |E_1|^2}
\nonumber \\
&=&
{\omega {\cal A}_H \over \pi} 
{(x_+^2-x_-^2)^{(\sqrt{1-s}-1)} \over 
{\Gamma(1+\sqrt{1-s})\Gamma(\sqrt{1-s})}}
\left \vert 
{{\Gamma({{1+\sqrt{1-s}} \over 2} - i { \omega \over 4 \pi T_L})
  \Gamma({{1+\sqrt{1-s}} \over 2} - i { \omega \over 4 \pi T_R})}
\over
  \Gamma(1 -i {\omega \over 2 \pi T_H})}
\right \vert^2,
\label{cross} 
\end{eqnarray} 
where left and right temperatures are defined by
\begin{equation}
{1 \over T_{L/R}} = {1 \over T_H} \left ( 1 \pm {r_- \over r_+} \right ).
\label{temperature}
\end{equation}
The absorption corss-section is given by 
$\sigma_{\rm abs} = {\cal A} / \omega$ in three dimensions.
Finally the decay rate is calculated as 
\begin{eqnarray}
\Gamma_{\rm fixed} = { \sigma_{\rm abs} \over {e^{\omega \over T_H} -1 }}
&=& {(x_+^2-x_-^2)^{\sqrt{1-s}} \over \pi \omega} 
{e^{-\omega/T_H} \over 
{\Gamma(1+\sqrt{1-s})\Gamma(\sqrt{1-s})}} 
\nonumber \\
&&~~~~\times
\left \vert 
{\Gamma({{1+\sqrt{1-s}} \over 2} - i { \omega \over 4 \pi T_L})
  \Gamma({{1+\sqrt{1-s}} \over 2} - i { \omega \over 4 \pi T_R})}
\right \vert^2.
\label{decay_rate}
\end{eqnarray}
This is the key result and is originated from (\ref{eq_decoupled})(
especially, the last term).
In the $s \to 0$ limit, (\ref{decay_rate}) recovers the decay rate for 
the minimally coupled scalar\cite{Bir97} 
\begin{equation}
\Gamma_{\rm min} 
= { \sigma_{\rm abs}^{\rm min} \over {e^{\omega \over T_H} -1 }}
= {{\pi l^2 \omega} \over 
 {(e^{\omega \over 2 T_R} -1) 
  (e^{\omega \over 2 T_L} -1)}}.
\label{decay_min}
\end{equation}
Note that (\ref{decay_min}) was derived from (\ref{eq_decoupled}) without 
the last term.
It is pointed out that the dilaton as a fixed scalar is only physically 
propagating field in the BTZ back ground.  
In the limit of $s \to -8$, our result (\ref{decay_rate}) recovers 
the same result for the dilaton as in Ref\cite{Teo98}. 
Finally we comment on the parameter $s$.  We introduce 
$s=-8 -\epsilon$ with the small parameter $\epsilon$ in 
Eq.(\ref{eq_far}).  This is so because $E_2$ in (\ref{aout}) 
has a pole for integral $s$.  Hence it is convenient 
to keep $s$ near an integer value during the calculation 
and make it integer at the end.

\section*{Acknowledgement}
This work was supported in part by the Basic Science Research Institute 
Program, Minstry of Education, Project NOs. BSRI-97-2441 and 
BSRI-97-2413.

\end{document}